
\input phyzzx
\input epsf
\FRONTPAGE
\line{\hfill BROWN-HET-1018}
\line{\hfill September 1995}
\bigskip
\titlestyle{IMPLEMENTING MARKOV'S LIMITING CURVATURE HYPOTHESIS
\foot{Contribution to the proceedings of the 6'th Quantum Gravity Seminar,
Moscow, June 1995, ed. V. Berezin et al. (World Scientific, Singapore, 1996).}}
\medskip
\author{Robert H. Brandenberger}
\centerline{{\it Department of Physics}}
\centerline{{\it Brown University, Providence, RI 02912, USA}}

\medskip
\abstract
The effective action for gravity at high curvatures is likely to contain higher
derivative terms. These corrections may have profound consequences for the
singularity structure of space-time and for early Universe cosmology. In this
contribution, recent work is reviewed which demonstrates that it is possible to
construct a class of effective gravitational actions for which all solutions
with sufficient symmetries have limited curvature and are nonsingular. Near the
limiting curvature, the coupling between matter and gravity goes to zero and in
this sense the theory is asymptotically free.
\endpage
\chapter{Introduction}

In 1982, Academician M. A. Markov, to whose memory this meeting is dedicated,
postulated$^1$ the existence of a limiting density for all forms of matter, a
density at which the microphysical differences between various types of matter
were assumed to disappear. For homogeneous and isotropic cosmologies, this
assumption would lead to a de Sitter phase at densities approaching the
limiting density. A slightly modified hypothesis, the limiting curvature
hypothesis$^2$, was put forwards a few years later. Here it is postulated that
for any solution of the gravitational equations of motion, in particular for a
black hole, the curvature remains bounded by a fundamental ``maximal
curvature", whose value we should expect to be given by the Planck scale. In
this case, the inside of a black hole would have to differ from the usual
Schwarschild solution. One possibility$^{2,3}$ is that it becomes a de Sitter
space-time throat leading to a different universe. This possibility has
recently also been suggested$^4$ in the context of string theory.

In this contribution, we will summarize some recent work$^{5-8}$ attempting to
implement the limiting curvature hypothesis.
We have
constructed an effective action for gravity in which all solutions
with sufficient symmetry are nonsingular.  The theory is a higher
derivative modification of the Einstein action, and is obtained by
a constructive procedure well motivated in analogy with the analysis
of point particle motion in special relativity.  The resulting theory
is asymptotically free in a sense which will be specified below.

The inclusion of higher derivative gravity terms in the fundamental Lagrangian
when studying the evolution of the space-time metric $g_{\mu\nu}$ at high
curvatures is well motivated, since it expected that
Planck scale physics will generate such types of correction terms to the
Einstein action.
This can be seen by considering the effective action obtained by
integrating out quantum matter fields in the presence of a dynamical
metric, by calculating first order perturbative quantum gravity
effects, or by studying the low energy effective action of a Planck
scale unified theory such as string theory.

The question we wish to address in this work is whether it is
possible to construct a class of effective actions for gravity which
have improved singularity properties
with the constraint that they give the correct low curvature limit. It is also
interesting to explore the implications of such models for early Universe
cosmology, in particular in connection with the possible occurrence of a period
of inflation.

A possible objection to our approach is that near a singularity
quantum effects will be important and therefore a classical analysis is
doomed to fail.  This argument is correct in the usual picture in
which at high curvatures there are large fluctuations and space-time
becomes more like a ``quantum foam."  However, in our theory, at high
curvature space-time becomes highly regular and thus a classical
analysis is self-consistent.  The property of asymptotic
freedom is essential in order to reach this conclusion.

Our aim is to construct a theory with the property that the metric
$g_{\mu\nu}$ approaches the de Sitter metric $g_{\mu\nu}^{DS}$, a
metric with maximal symmetry which admits a geodesically complete and
nonsingular extension, as the curvature $R$ approaches the Planck
value $R_{pl}$.  Here, $R$ stands for any curvature invariant.
Naturally, from our classical considerations, $R_{pl}$ is a free
parameter.  However, if our theory is connected with Planck scale
physics, we expect $R_{pl}$ to be set by the Planck scale.

\epsfxsize=6in \epsfbox{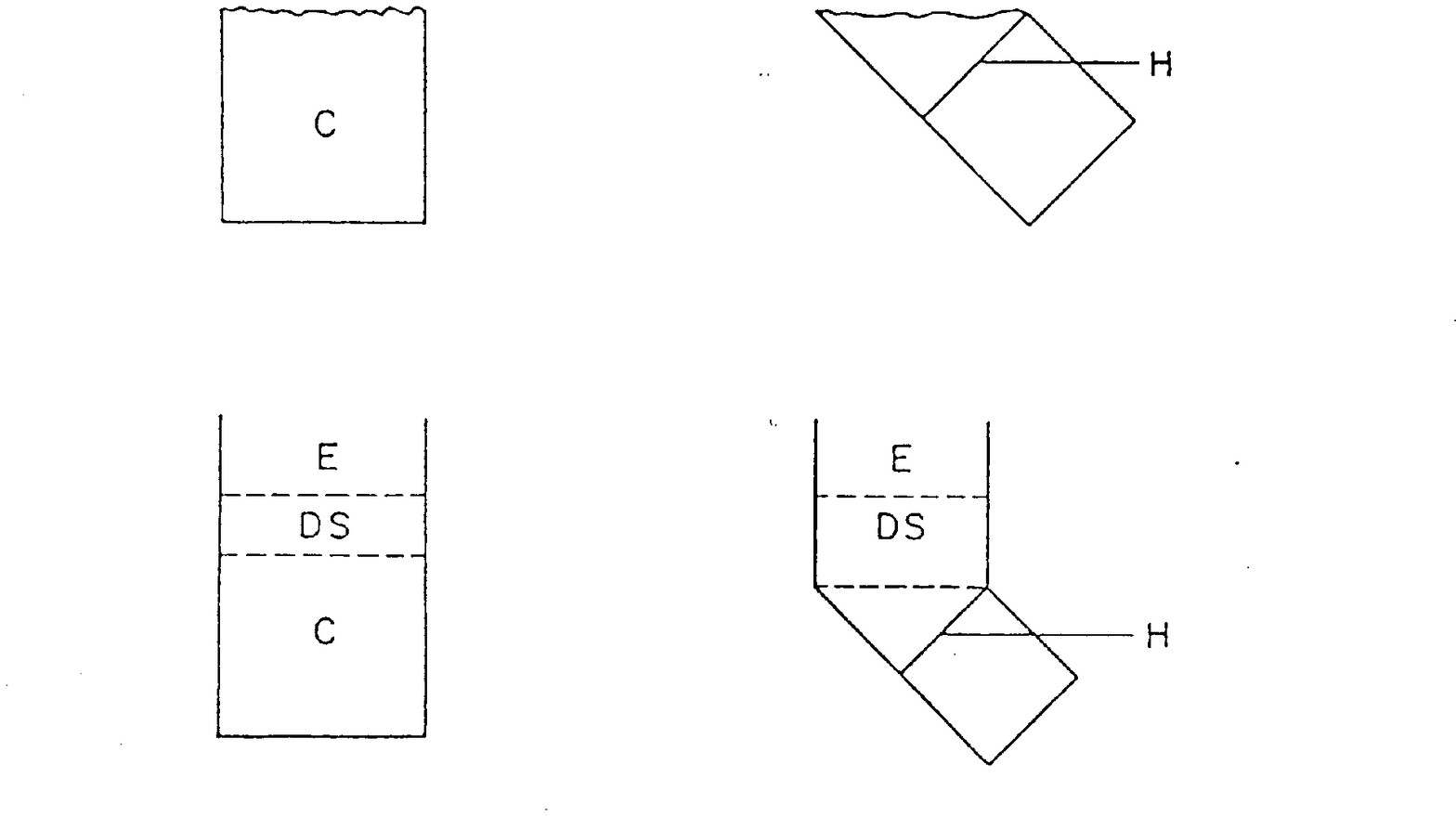}
{\baselineskip=13pt
\noindent{\bf Figure 1:} Penrose diagrams for collapsing Universe (left) and
black hole (right) in Einstein's theory (top) and in the nonsingular Universe
(bottom). C, E, DS and H stand for contracting phase, expanding phase, de
Sitter phase and horizon, respectively, and wavy lines indicate singularities.}

If successful, the above construction will have some very appealing
consequences.  Consider, for example, a collapsing spatially
homogeneous Universe.  According to Einstein's theory, this Universe
will collapse in finite proper time to a final ``big crunch" singularity (top
left Penrose diagram of Figure 1).
In our theory, however, the Universe will approach a de Sitter model as
the curvature increases.  If the
Universe is closed, there will be a de Sitter bounce followed by
re-expansion (bottom left Penrose diagram in Figure 1).  Similarly, in our
theory spherically
symmetric vacuum  solutions would be nonsingular, i.e., black holes
would have no singularities in their centers.  The structure of a
large black hole would be unchanged compared to what is predicted by
Einstein's theory (top right, Figure 1) outside and even slightly inside the
horizon, since
all curvature
invariants are small in those regions.  However, for $r \rightarrow 0$
(where $r$ is the radial Schwarzschild coordinate), the solution
changes and approaches a de Sitter solution (bottom right, Figure 1).  This
would have interesting consequences for the black hole information
loss problem.

To motivate our effective action construction, we turn to a well known
analogy, point particle motion in the theory of special relativity.

\chapter{An Analogy}

 The transition from the Newtonian theory of point particle motion to
the special relativistic theory transforms a theory with no bound on
the velocity into one in which there is a limiting velocity, the speed
of light $c$ (in the following we use units in which $\hbar = c = 1$).
This transition can be obtained$^5$ by starting with the action of a
point particle with world line $x(t)$:
$$
S_{\rm old} = \int dt {1\over 2} \dot x^2 \, , \eqno\eq
$$
and adding$^9$ a Lagrange multiplier which couples to $\dot
x^2$, the quantity to be made finite, and which has a potential
$V(\varphi)$:
$$
S_{\rm new} = \int dt \left[ {1\over 2} \dot x^2 + \varphi \dot x^2 -
V (\varphi) \right] \, .\eqno\eq
$$
{}From the constraint equation
$$
\dot x^2 = {\partial V\over{\partial \varphi}} \, , \eqno\eq
$$
it follows that $\dot x^2$ is limited provided $V(\varphi)$ increases
no faster than linearly in $\varphi$ for large $|\varphi|$.  The small
$\varphi$ asymptotics of $V(\varphi)$ is determined by demanding that
at low velocities the correct Newtonian limit results:
$$
\eqalign{V (\varphi) \sim \varphi^2 \> & {\rm as} \> |\varphi|
\rightarrow 0 \, , \cr
V (\varphi) \sim \varphi \> & {\rm as} \> |\varphi| \rightarrow \infty
\, . } \eqno\eq
$$
Choosing the simple interpolating potential
$$
V (\varphi) = {2 \varphi^2\over{1 + 2 \varphi}} \, , \eqno\eq
$$
the Lagrange multiplier can be integrated out, resulting in the well-known
action
$$
S_{\rm new} = {1\over 2} \int dt \sqrt{1 - \dot x^2} \eqno\eq
$$
for point particle motion in special relativity.

\chapter{Construction}

 Our procedure for obtaining a nonsingular Universe theory$^5$ is based
on generalizing the above Lagrange multiplier construction to gravity.
Starting from the Einstein action, we can introduce a Lagrange
multiplier $\varphi_1$ coupled to the Ricci scalar $R$ to obtain a
theory with limited $R$:
$$
S = \int d^4 x \sqrt{-g} (R + \varphi_1 \, R + V_1 (\varphi_1) ) \, ,
\eqno\eq
$$
where the potential $V_1 (\varphi_1)$ satisfies the asymptotic
conditions (2.4).

However, this action is insufficient to obtain a nonsingular gravity
theory.  For example, singular solutions of the Einstein equations
with $R=0$ are not effected at all.  The minimal requirements for a
nonsingular theory is that \underbar{all} curvature invariants remain
bounded and the space-time manifold is geodesically complete.
Implementing Markov's Limiting Curvature Hypothesis$^{1,2}$, these conditions
can be reduced to more manageable ones.  First, we choose one
curvature invariant $I_1 (g_{\mu\nu})$ and demand that it be
explicitely bounded, i.e., $|I_1| < I_1^{pl}$, where $I_1^{pl}$ is the
Planck scale value of $I_1$.  In a second step, we demand that as $I_1
(g_{\mu\nu})$ approaches $I_1^{pl}$, the metric $g_{\mu\nu}$ approach
the de Sitter metric $g^{DS}_{\mu\nu}$, a definite nonsingular metric
with maximal symmetry.  In this case, all curvature invariants are
automatically bounded (they approach their de Sitter values), and the
space-time can be extended to be geodesically complete.

Our approach is to implement the second step of the above procedure by
another Lagrange multiplier construction$^5$.  We look for a curvature
invariant $I_2 (g_{\mu\nu})$ with the property that
$$
I_2 (g_{\mu\nu}) = 0 \>\> \Leftrightarrow \>\> g_{\mu\nu} =
g^{DS}_{\mu\nu} \, , \eqno\eq
$$
introduce a second Lagrange multiplier field $\varphi_2$ which couples
to $I_2$ and choose a potential $V_2 (\varphi_2)$ which forces $I_2$
to zero at large $|\varphi_2|$:
$$
S = \int d^4  x \sqrt{-g} [ R + \varphi_1 I_1 + V_1 (\varphi_1) +
\varphi_2 I_2 + V_2 (\varphi_2) ] \, , \eqno\eq
$$
with asymptotic conditions (2.4) for $V_1 (\varphi_1)$ and conditions
$$
\eqalign{V_2 (\varphi_2) & \sim {\rm const} \>\> {\rm as} \> |
\varphi_2 | \rightarrow \infty \cr
V_2 (\varphi_2) & \sim \varphi^2_2 \>\> {\rm as} \> |\varphi_2 |
\rightarrow 0 \, ,} \eqno\eq
$$
for $V_2 (\varphi_2)$.  The first constraint forces $I_2$ to zero, the
second is required in order to obtain the correct low curvature limit.

These general conditions are reasonable, but not sufficient in order
to obtain a nonsingular theory.  It must still be shown that all
solutions are well behaved, i.e., that they asymptotically reach the
regions $|\varphi_2| \rightarrow \infty$ of phase space (or that
they can be controlled in some other way).  This must be done for a
specific realization of the above general construction.

\chapter{Specific Model}

At the moment we are only able to find an invariant $I_2$ which
singles out de Sitter space by demanding $I_2 = 0$ provided we assume
that the metric has special symmetries.  The choice
$$
I_2 = (4  R_{\mu\nu} R^{\mu\nu} - R^2 + C^2)^{1/2} \, , \eqno\eq
$$
singles out the de Sitter metric among all homogeneous and isotropic
metrics (in which case adding $C^2$, the Weyl tensor square, is
superfluous), all homogeneous and anisotropic metrics, and all
radially symmetric metrics.

We choose the action$^{5,6}$
$$
S = \int d^4 x \sqrt{-g} \left[ R + \varphi_1 R - (\varphi_2 +
{3\over{\sqrt{2}}} \varphi_1) I_2^{1/2} + V_1 (\varphi_1) + V_2
(\varphi_2) \right] \eqno\eq
$$
with
$$
V_1 (\varphi_1) = 12 \, H^2_0 {\varphi^2_1\over{1 + \varphi_1}} \left( 1
- {\ln (1 + \varphi_1)\over{1 + \varphi_1}} \right) \eqno\eq
$$
$$
V_2 (\varphi_2) = - 2 \sqrt{3} \, H^2_0 \, {\varphi^2_2\over{1 +
\varphi^2_2}} \, . \eqno\eq
$$

The general equations of motion resulting from this action are quite
messy.  However, when restricted to homogeneous and isotropic metrics
of the form
$$
ds^2 = dt^2 - a (t)^2 (dx^2 + dy^2 + dz^2) \, , \eqno\eq
$$
the equations are fairly simple.  With $H = \dot a / a$, the two
$\varphi_1$ and $\varphi_2$ constraint equations are
$$
H^2 = {1\over{12}} V^\prime_1 \eqno\eq
$$
$$
\dot H = - {1\over{2\sqrt{3} }} V^\prime_2 \, , \eqno\eq
$$
and the dynamical $g_{00}$ equation becomes
$$
3 (1 - 2 \varphi_1) H^2 + {1\over 2} (V_1 + V_2) = \sqrt{3} H (\dot
\varphi_2 + 3 H \varphi_2) \, . \eqno\eq
$$
The phase space of all vacuum configurations is the half plane $\{
(\varphi_1 \geq 0, \, \varphi_2) \}$.  Equations (4.6) and (4.7)
can be used to express $H$ and $\dot H$ in terms of $\varphi_1$ and
$\varphi_2$.  The remaining dynamical equation (4.8) can then be recast
as
$$
{d \varphi_2\over{d \varphi_1}} = - {V_1^{\prime\prime}\over{4
V^\prime_2}} \, \left[ - \sqrt{3} \varphi_2 + (1 - 2\varphi_1) -
{2\over{V^\prime_1}} (V_1 + V_2) \right] \, . \eqno\eq
$$
The solutions can be studied analytically in the asymptotic regions
and numerically throughout the entire phase space.

The resulting phase diagram of vacuum solutions is sketched in Fig. 2
(for numerical results, see Ref. 6).  The point $(\varphi_1, \,
\varphi_2) = (0,0)$ corresponds to Minkowski space-time $M^4$, the
regions $|\varphi_2 | \rightarrow \infty$ to de Sitter space.  As
shown, all solutions either are periodic about $M^4$ or else they
asymptotically approach de Sitter space.  Hence, all solutions are
nonsingular.  This conclusion remains unchanged if we add spatial
curvature to the model.

\epsfxsize=6in \epsfbox{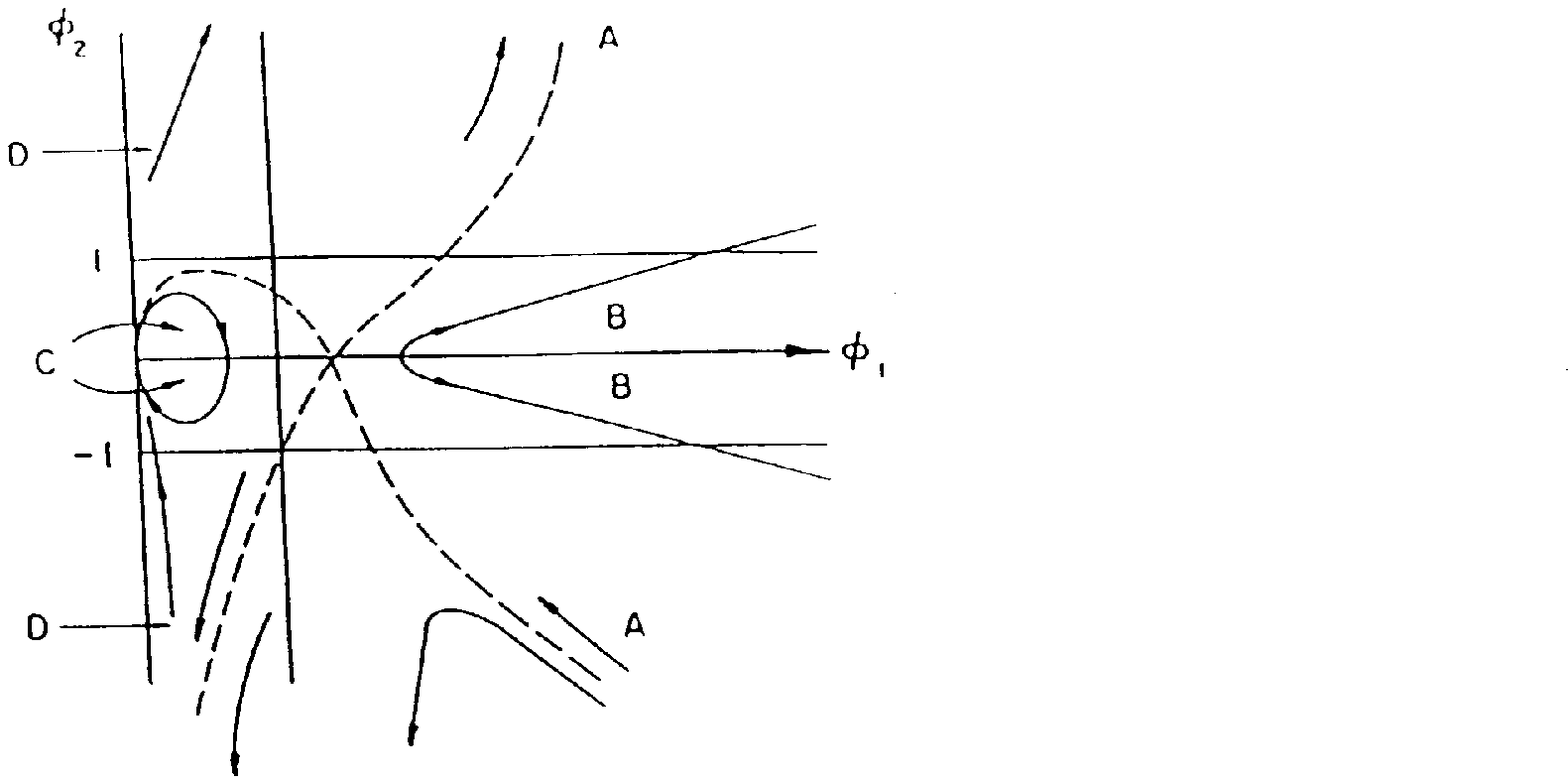}
{\baselineskip=13pt
\noindent{\bf Figure 2:} Phase diagram of the homogeneous and isotropic
solutions of the nonsingular Universe. The asymptotic regions are labelled by
A, B, C and D, flow lines are indicated by arrows.}

One of the most interesting properties of our theory is asymptotic
freedom$^6$, i.e., the coupling between matter and gravity goes to
zero at high curvatures.  It is easy to add matter (e.g., dust or
radiation) to our model by taking the combined action
$$
S = S_g + S_m \, , \eqno\eq
$$
where $S_g$ is the gravity action previously discussed, and $S_m$ is
the usual matter action in an external background space-time metric.

We find$^6$ that in the asymptotic de Sitter regions, the trajectories of
the solutions in the $(\varphi_1, \, \varphi_2)$ plane are unchanged
by adding matter.  This applies, for example, in a phase of de Sitter
contraction when the matter energy density is increasing exponentially
but does not affect the metric.  The physical reason for asymptotic
freedom is obvious: in the asymptotic regions of phase space, the
space-time curvature approaches its maximal value and thus cannot be
changed even by adding an arbitrary high matter energy density.

Naturally, the phase space trajectories near $(\varphi_1, \,
\varphi_2) = (0,0)$ are strongly effected by adding matter.  In
particular, $M^4$ ceases to be a stable fixed point of the evolution
equations.

\chapter{Two Dimensional Results}

 The low energy effective actions for the space-time metric in 4
dimensions which come from string theory are only known
perturbatively.  They contain higher derivative terms, but not if the
exact same form as the ones used in our construction.  The connection
between our limiting curvature construction and string theory-motivated
effective actions is more apparent in two
 space-time dimensions$^{7,8}$.

The most general renormalizable Lagrangian for string-induced dilaton
gravity is
$$
{\cal L} = \sqrt{-g} [ D(\varphi) R + G (\varphi) (\nabla \varphi)^2 +
H (\varphi) ] \, , \eqno\eq
$$
where $\varphi (x,t)$ is the dilaton.  In two space-time dimensions,
the kinetic term for $\varphi$ can be eliminated, resulting in a
Lagrangian (in terms of rescaled fields) of the form
$$
{\cal L} = \sqrt{-g} [ D(\varphi) R + V (\varphi) ] \, . \eqno\eq
$$

We can now apply the limiting curvature construction to find classes
of potentials for which the theory has nonsingular black hole$^7$ and
cosmological$^8$ solutions.  In the following, we discuss the
nonsingular two-dimensional black hole. For other discussions of nonsingular
two-dimensional black holes and cosmological solutions see Refs. 10 and 11,
respectively.

To simplify the algebra, the dilaton is redefined such that
$$
D (\varphi) = {1\over \varphi} \, . \eqno\eq
$$
The most general static metric can be written as
$$
ds^2 = f (r) dt^2 - g (r) dr^2 \eqno\eq
$$
and the gauge choice
$$
g (r) = f (r)^{-1} \eqno\eq
$$
is always possible.  The variational equations are
$$
f^\prime = - V (\varphi) {\varphi^2\over \varphi^\prime} \, , \eqno\eq
$$
$$
\left( {\varphi^\prime\over \varphi^2} \right)^\prime = 0 \eqno\eq
$$
and
$$
\varphi^{-2} R = {\partial V\over{\partial \varphi}} \, , \eqno\eq
$$
where a prime denotes the derivative with respect to $r$.

Equation (5.7) can be integrated to find (after rescaling $r$)
$$
\varphi = {1\over{Ar}} \, . \eqno\eq
$$
To give the correct large $r$ behavior for the metric, we need to
impose that
$$
f (r) \rightarrow 1 - {2m\over r} \>\>\> {\rm as} \> r \rightarrow
\infty \, . \eqno\eq
$$
{}From (5.6) this leads to the asymptotic condition
$$
V (\varphi) \rightarrow 2 m A^3 \varphi^2 \>\>\> {\rm as} \> \varphi
\rightarrow 0 \, . \eqno\eq
$$
The limiting curvature hypothesis requires that $R$ be bounded as
$\varphi \rightarrow \infty$.  From (5.8) this implies
$$
V (\varphi) \rightarrow {2\over{\ell^2 \varphi}} \>\>\> {\rm as} \>
\varphi \rightarrow \infty \, , \eqno\eq
$$
where $\ell$ is a constant which determines the limiting curvature.
As an interpolating potential we can choose
$$
V (\varphi) = {2 m A^3 \varphi^2\over{1+ m A^3 \ell^2 \varphi^3}} \, ,
\eqno\eq
$$
which allows (5.6) to be integrated explicitly$^7$ to obtain $f(r)$.

The resulting metric coefficient $f(r)$ describes a nonsingular black
hole with a single horizon at $r \simeq 2m$.  The metric is
indistinguishable from the usual Schwarzschild metric until far inside
of the horizon, where our $f(r)$ remains regular and obtains vanishing
derivative at $r = 0$, which allows for a geodesically complete
extension of the manifold.

\chapter{Discussion}

We have shown that a class of higher derivative extensions of the
Einstein theory exist for which many interesting solutions are
nonsingular.  Our class of models is very special.  Most higher
derivative theories of gravity have, in fact, much worse singularity
properties than the Einstein theory.  What is special about our class
of theories is that they are obtained using a well motivated Lagrange
multiplier construction which implements the limiting curvature
hypothesis.  We have shown that
\item{\rm i)} all homogeneous and isotropic solutions are
nonsingular$^{5,6}$
\item{\rm ii)} the two-dimensional black holes are nonsingular$^7$
\item{\rm iii)} nonsingular two-dimensional cosmologies exist$^8$.

\noindent
We also have evidence that four-dimensional black holes and
anisotropic homogeneous cosmologies are nonsingular$^{12}$.

By construction, all solutions are de Sitter at high curvature.  Thus,
the theories automatically have a period of inflation (driven by the
gravity sector in analogy to Starobinsky inflation$^{13}$) in the
early Universe.

A very important property of our theories is asymptotic freedom.  This
means that the coupling between matter and gravity goes to zero at
high curvature, and might lead to an automatic suppression mechanism
for scalar fluctuations.

In two space-time dimensions, there is a close connection between
dilaton gravity and our construction.  In four dimensions, the
connection between fundamental physics and our class of effective
actions remains to be explored. A promising direction for future research
appears to be an exploration of the connection between the nonsingular
cosmology described here and the ``pre-big-bang" scenario$^{14}$ which is based
on string-inspired dilaton gravity. Using our implementation of the Limiting
Curvature Hypothesis it might be possible to resolve the ``graceful exit
problem"$^{15}$ of dilaton gravity.

More immediately, however, there are many important problems concerning the
construction proposed here which remain to be resolved. In particular, does the
theory remain well behaved when allowing for space-times without the special
symmetries which we have assumed? What is the behavior of inhomogeneities at
the linearized level? Does asymptotic freedom of the de Sitter phase have an
effect on the magnitude of the density fluctuations produced during inflation?
At first sight, no fundamental obstacles have appeared. However, the actual
computations appear extremely tedious as a consequence of the higher derivative
terms which appear in the action. Nevertheless, the potential benefits of our
scenario make these computations well worth while.

\ack

I am grateful to Professors V. Berezin and V. Rubakov for inviting me to write
this contribution. I also wish
to thank my collaborators Richhild Moessner, Masoud Mohazzab, Andrew
Sornborger,  Mark Trodden and in particular Slava Mukhanov for the joy of
collaboration.
This work is supported in part by the US Department of Energy under
Grant DE-FG0291ER40688, Task A.

\REF\one{M. Markov, {\it Pis'ma Zh. Eksp. Theor. Fiz.} {\bf 36}, 214
(1982); \nextline
M. Markov, {\it Pis'ma Zh. Eksp. Theor. Fiz.} {\bf 46}, 342 (1987).}
\REF\two{V. Frolov, M. Markov and V. Mukhanov, {\it Phys. Lett.} {\bf B216},
272 (1989);\nextline
V. Frolov, M. Markov and V. Mukhanov, {\it Phys. Rev.} {\bf D41}, 383
(1990).}
\REF\three{D. Morgan, {\it Phys. Rev.} {\bf D43}, 3144 (1991); \nextline
I. Dymnikova, {\it Gen. Rel. Grav.} {\bf 24}, 235 (1992).}
\REF\four{E. Martinec, {\it Class. Quant. Grav.} {\bf 12}, 941 (1995).}
\REF\five{V. Mukhanov and R. Brandenberger, {\it Phys. Rev. Lett.} {\bf
68}, 1969 (1992).}
\REF\six{R. Brandenberger, V. Mukhanov and A. Sornborger, {\it Phys.
Rev.} {\bf D48}, 1629 (1993).}
\REF\seven{M. Trodden, V. Mukhanov and R. Brandenberger, {\it Phys.
Lett.} {\bf B316}, 483 (1993).}
\REF\eight{R. Moessner and M. Trodden, {\it Phys. Rev.} {\bf D51}, 2801
(1995).}
\REF\nine{B. Altshuler, {\it Class. Quant. Grav.} {\bf 7}, 189
(1990).}
\REF\ten{R. Mann, S. Morsink, A. Sikkema and T. Steele, {\it Phys. Rev.} {\bf
D43}, 3948 (1991); \nextline
R. Mann, {\it Gen. Rel. Grav.} {\bf 24}, 433 (1992); \nextline
D. Christensen and R. Mann, {\it Class. Quant. Grav.} {\bf 9}, 1769 (1992).}
\REF\eleven{T. Mishima and A. Nakamichi, {\it Prog. Theor. Phys. Suppl.} {\bf
114}, 207 (1993); \nextline
M. Yoshimura, {\it Phys. Rev.} {\bf D47}, 5389 (1993); \nextline
K. Chan and R. Mann, {\it Class. Quant. Grav.} {\bf 10} 913 (1993); \nextline
M. Osorio and M. Vazquez-Mozo, {\it Mod. Phys. Lett.} {\bf A8}, 3111 (1993);
\nextline
M. Osorio and M. Vazquez-Mozo, {\it Mod. Phys. Lett.} {\bf A8}, 3215 (1993).}
\REF\twelve{R. Brandenberger, M. Mohazzab, V. Mukhanov, A. Sornborger and
M. Trodden, in preparation (1995).}
\REF\thirteen{A. Starobinsky, {\it Phys. Lett.} {\bf B91}, 99 (1980).}
\REF\fourteen{M. Gasperini and G. Veneziano, {\it Phys. Lett.} {\bf B277}, 256
(1992);\nextline
M. Gasperini and G. Veneziano, {\it Astropart. Phys.} {\bf 1}, 317 (1993);
\nextline
J. Levin, {\it Phys. Rev.}, {\bf D51}, 1536 (1995).}
\REF\fifteen{R. Brustein and G. Veneziano, {\it Phys. Lett.} {\bf B329}, 429
(1994);\nextline
N. Kaloper, R. Madden and K. Olive, `Towards a Singularity-Free Inflationary
Universe', hep-th/9506027 (1995).}
\refout
\bye